\documentstyle[12pt,epsf]{article}
\def\sublabel#1#2{\@bsphack\if@filesw {\let\thepage\relax
   \def\protect{\noexpand\noexpand\noexpand}%
   \edef\@tempa{\write\@auxout{\string
      \newlabel{#1}{{\@currentlabel#2}{\thepage}}}}%
   \expandafter}\@tempa
   \if@nobreak \ifvmode\nobreak\fi\fi\fi\@esphack}

\newcommand{\twographs}[2]{%
   \unitlength=1in
   \begin{center}
     \begin{picture}(6,2.2)
       \put(0,0){\labgraph{a}{#1}}
       \put(3,0){\labgraph{b}{#2}}
     \end{picture}
   \end{center}}

\newcommand{\labgraph}[2]{%
   \begin{picture}(3,2)
       \put(0,0){\makebox(3,2)%
{\centering\epsfxsize=2.6in\leavevmode\epsffile{#2.ps}}}
       \put(0,1){\makebox(0,0)[l]{(#1)}}
   \end{picture}}

\advance \topmargin by -\headheight
\advance \topmargin by -\headsep

\evensidemargin \oddsidemargin
 \advance\hoffset by -3mm  
 \advance\voffset by  8mm  

\begin{document}

\baselineskip 12pt

\newcommand{\sheptitle}
{Renormalization Group Naturalness\\ \medskip of GUT Higgs Potentials}

\newcommand{\shepauthor}
{B. C. Allanach$^1_a$, G. Amelino-Camelia$^2_b$, O. Philipsen$^3_c$, 
O. Pisanti$^{4,5}_d$, and L. Rosa$^4_e$} 

\newcommand{\shepaddress}
{1. Department of Particle Physics, Rutherford Appleton Laboratory,\\
Chilton, Didcot, Oxon, OX11 OQX, UK \\ 
\smallskip
2. Institut de Physique, Universit\'e de Neuch\^atel,\\
rue Breguet 1, CH-2000 Neuch\^atel, Switzerland \\
\smallskip
3. Institut f\"ur Theoretische Physik, Universit\"at Heidelberg,\\
Philosophenweg 16, 69120 Heidelberg, Germany \\
\smallskip
4. Dipartimento di Scienze Fisiche, Universit\`a di Napoli, and INFN, Sez.
di Napoli,\\
Mostra d'Oltremare, Pad.19, I-80125 Napoli, Italy \\
\smallskip
5. Kellogg Radiation Laboratory,
California Institute of Technology,\\
Pasadena, CA 91125, USA} 

\newcommand{\shepabstract}
{We analyze the symmetry-breaking patterns of grand unified
theories from the point 
of view of a recently-proposed criterion 
of renormalization-group naturalness.
We perform the analysis on simple 
non-SUSY SU(5) and SO(10)
and SUSY SU(5) GUTs.
We find that the naturalness criterion can favor
spontaneous-symmetry-breaking in the direction
of the smallest of the maximal little groups.
Some differences between theories with and without
supersymmetry are also emphasized.}

\begin{titlepage}
\begin{flushright}
{\tt hep-ph/9804313}\\
NEIP-98-001\\
HD-THEP-98-14\\
DSF-1/98\\
MAP-221
\end{flushright}
\vspace{.4in}
\begin{center}
{\large{\bf \sheptitle}}
\bigskip \\ \shepauthor \\ \mbox{} \\ {\it \shepaddress} \\
\vspace{.3in}
{\bf Abstract} \smallskip \end{center} \setcounter{page}{0}
\shepabstract
\bigskip
\bigskip
\vfill \noindent
{\tt a) bca@hep.phys.soton.ac.uk \\
b) giovanni.amelino-camelia@cern.ch \\
c) o.philipsen@thphys.uni-heidelberg.de\\
d) pisanti@na.infn.it\\
e) rosa@na.infn.it}
\end{titlepage}

\section{Introduction}

A large number of grand unified theories (GUTs) has
already been discussed in the literature,
and, assuming that the GUT idea
is successful, we are confronted 
with the fact that 
the constraints imposed by the low-energy data 
are not sufficient to select the correct GUT.
Certain additional constraints on the
structure of a GUT emerge if one
requires that it be a plausible candidate
as low-energy effective description of a more fundamental
theory  (for example including gravity).
In particular, in Ref.~\cite{gacrus} it was observed
that one could study the RG equations describing the
running of the parameters of the Higgs potential
between $M^*$, denoting the scale
(possibly given by the Planck
scale $M_P \! \sim \! 10^{19} GeV$)
where the GUT emerges as low-energy effective theory,
and the GUT scale $M_X$ (the scale
where the low energy couplings unify).
This could establish whether or not
a given pattern of SSB (spontaneous symmetry breaking)
can be naturally obtained.
In this context, we define a symmetry breaking direction 
to be natural if it
corresponds to a large volume of parameter space at $M^*$.
For example,
if a strongly attractive fixed point was found within a region 
of the space of parameters of the Higgs
potential that corresponds to a certain SSB pattern,
one would then expect that 
in running toward the infra-red direction,
the Higgs parameters would approach their fixed point values
(forcing the corresponding SSB pattern) 
quite independently of the input parameters at the scale $M^*$.

Importantly, this {\it RG naturalness} criterion
concerns one of the non-predictive aspects 
of GUTs, {\it i.e.} the SSB pattern. Even after
selecting the matter (Higgs) content, most GUTs may break
in several different ways depending on which vacuum expectation
values are acquired by the Higgs fields. If, as ordinarily assumed,
the direction of SSB is determined by the global minimum of the
Higgs potential, the symmetry-breaking pattern of a GUT
only depends upon the input parameters in
the Higgs potential of the model. 
While we are discussing here the first step of GUT
SSB, similar RG naturalness
considerations could  of course be applied to other steps of SSB.

In Ref.~\cite{aap} the criterion
of RG naturalness
has already
been applied to the study of the first step of SSB
of a SUSY (supersymmetric) $SU(5)$ GUT.
Interestingly, an infra-red fixed point was identified
analytically, and it was found to be located at the boundary
between the region of Higgs parameter space
corresponding to unbroken SU(5) and the region corresponding
to the breaking of SU(5) to the Standard Model gauge group
$G_{SM}$~$\equiv$~$SU(3) \otimes SU(2) \otimes U(1)$.
Also motivated by the fact that the presence of a fixed point
at the boundary between two SSB regions 
may not be typical,
we intend to illustrate a wider range of possible
scenarios for the outcome of a RG naturalness analysis.
We do this by studying the first step of SSB
of two additional examples, i.e. non-SUSY $SU(5)$ 
and $SO(10)$ GUTs.
In order to discuss the role
that SUSY might play in RG naturalness analyses,
we also briefly review the results of Ref.~\cite{aap}.

In the remainder of this introduction we mention a few other
RG approaches which have been used in the study of GUTs,
and we comment on the relation between these approaches
and the type of approach advocated here.
Let us start by mentioning those studies 
(see, {\it e.g.}, Ref.~\cite{ross} 
and references therein)  that
have established how predictions for the low-energy 
values of certain quantities can be obtained from the infra-red structure 
of the relevant RG equations. 
While these studies do not always assume
grand unification, the fact that the values 
of (low-energy) parameters can be derived from the structure of the RG
equations
is encouraging for our attempt 
to derive 
the SSB pattern in a similar way. 

RG techniques have also already 
been applied to the study of GUT-scale physics; notably
in the context of certain analyses of stability\cite{orafstab,MandR}.
While the actual formulae one ends up studying are in some cases 
the same, the emphasis in
stability analyses 
is quite different from the one of the present article, 
especially since
the stability characterizes a theory at the GUT scale, 
whereas here one is interested in the physics between 
the scale $M^*$ and the GUT scale.

Another example of RG ideas applied to GUT-scale physics
is given by the studies of finite GUTs\cite{piguetkazak}.
Some of the finite GUTs that correspond to an IR fixed point
of the RG equations for the Higgs parameters
might provide a good starting point for the construction of
an RG natural GUT;
however, dedicated analyses are necessary since the literature
on finite GUTs has often not considered SSB
and the parameters of the Higgs potential  in detail.

Finally, we should mention the idea 
of radiative breaking \cite{radbreak},
in which the symmetry-breaking term
is induced directly by the RG running,
rather than being included by hand in the Higgs potential
at a given scale.
This idea is based on an intuition
that is very close to the one behind 
the criterion of RG naturalness.
The two proposals primarily differ in that the former restricts 
the analysis to models in which the symmetry-breaking term
is directly induced by the RG running, whereas the latter
includes models in which a non-vanishing symmetry-breaking term
is already introduced at the scale $M^*$ as long as SSB occurs in the
direction favored by the RG running.

We now turn to explicit examples of GUT Higgs potentials.
The analysis reported in the following sections should also
illustrate the similarities 
and the differences between RG naturalness analyses and the above
approaches concerning the use of RG 
equations in the study of grand unification.

\section{A SUSY SU(5) model}

Let us start, as anticipated, with a brief review of the results obtained
in Ref.~\cite{aap}, where the criterion of RG naturalness
was applied to the study of the first step of SSB
of the minimal SUSY SU(5) GUT. This model
involves the Higgs fields of the 24-dimensional
irreducible representation (the adjoint), which are
responsible for the first step of SSB,
and it also includes $\underline{5} +
\overline{\underline{5}}$ Higgs, which are used in the second SSB step.
However, for simplicity in our analysis of the first SSB step we neglect
the effects of the $\underline{5} + \overline{\underline{5}}$ Higgs. 
We therefore limit our analysis to the potentials involving the
$\underline{24}$ Higgs. The superpotential is taken to be~\cite{su5} 
\begin{equation}
W = \lambda_1 \mbox{Tr}(\Sigma^3) + \mu \mbox{Tr} (\Sigma^2) \, ,
\label{Wsu5}
\end{equation}
where $\Sigma$ denotes the $24$-dimensional superfield multiplet. We assume
that SUSY breaking is explicit, via the ``soft'' SUSY-breaking terms in the
potential 
\begin{equation}
V_{soft} = \left[\frac{m_3}{6} \mbox{Tr} (\sigma^3) + m_2^2 \mbox{Tr}
(\sigma^2) +\frac{M}{2} \lambda \lambda + {\rm h.c.}\right]
+ m^2_{3/2} \mbox{Tr}(\sigma^{\dag}\sigma) \;,
\label{su5soft}
\end{equation}
where $\sigma$ represents the scalar component of $\Sigma$ and $\lambda$
denotes the SU(5) gaugino. The full Higgs potential relevant for the first
step of SSB can be written as 
\begin{eqnarray}
V = \left | {\partial W \over \partial \Sigma_i} \right |^2
+ V_{soft} + \mbox{D-terms}~.\label{pot}
\end{eqnarray}
Based on hierarchy arguments~\cite{hier} we expect $m_3, m_{3/2}, M \sim
1$~TeV and $m_2 \sim 10^{11}$~GeV, while $\mu$ is a GUT scale 
parameter\footnote{In
GUT-scale radiative-breaking scenarios one
considers the possibility $\mu \! = \! 0$, which is
stable under the one-loop RG 
equations. In the present work we shall ignore this
possibility. Its analysis would require a generalization of our study of
the Higgs potential, not relying on the simplifications we achieve by
assuming $|m_i/\mu| \! \ll \! 1$.} 
expected to be of order 10$^{16}$~GeV. 

It is convenient 
to consider the three independent combinations
of parameters
$\delta_2 \! \equiv \! m^2_2/\mu^2$, $\delta_3 \! \equiv \! m_3/\mu$ and
$\delta_{3/2} \! \equiv \! m_{3/2}/\mu$, which allow to rewrite 
the soft-breaking potential as~\cite{aap}
\begin{equation} \label{delv}
V_{soft}=\frac{8\mu^4}{27\lambda_1^2} \, b \, F \, ,
\end{equation}
where $b \! = \! 0$ in the minimum preserving 
the full $SU(5)$ invariance,
$b \! = \! 30$ for the $G_{SM}$-invariant minimum, 
$b \! = \! 20/9$ for the $SU(4) \otimes U(1)$-invariant minimum
and
\begin{equation} \label{newvsoft}
F \equiv 3\delta_2-\frac{1}{3\lambda_1}\delta_3 +\frac{3}{2}\delta_{3/2}^2
\,~.
\end{equation}
Hence, the value of $F$ at the
GUT scale $M_X$ determines the type of residual symmetry below $M_X$. 
If $F \! < \! 0$ the $G_{SM}$-invariant minimum is the lowest one,
while $SU(5)$ remains unbroken if $F \! > \! 0$. 

The one-loop RG equations may be easily derived~\cite{aap}
following the general prescriptions of Martin and Vaughn~\cite{MandV}, 
\begin{eqnarray}
16 \pi^2 \frac{d \lambda_1}{d t} &=& 3 \lambda_1 \left(\frac{189}{40}
\lambda_1^2 - 10 g^2\right) \label{su5RGEsa} \\ 
16 \pi^2 \frac{d \mu}{d t} &=& 2 \mu \left(\frac{189}{40} \lambda_1^2 - 10
g^2\right) \label{su5RGEsb} \\ 
16 \pi^2 \frac{d m_3}{d t} &=& 3 \left[m_3\left(\frac{189}{40} \lambda_1^2
- 10 g^2\right) +\frac{189}{20}\lambda_1^2 m_3 + 120 M\lambda_1 g^2 \right]
\label{su5RGEsc} \\ 
16 \pi^2 \frac{d m_2^2}{d t} &=& 2 \left[m_2^2 \left(\frac{189}{40}
\lambda_1^2 - 10 g^2\right) + \frac{63}{40}\lambda_1\mu m_3 + 20 M \mu g^2
\right] \label{su5RGEsd} \\ 
16\pi^2 \frac{d m^2_{3/2}}{d t} &=& \frac{567}{20}\lambda_1^2 m_{3/2}^2 +
\frac{21}{80} m_3^2 - 40 M^{\dag}Mg^2 \label{su5RGEse} \\ 
16 \pi^2 \frac{d g^2}{d t} &=& \beta g^4 \label{su5RGEsf} \\ 
16 \pi^2 \frac{d M}{d t} &=& \beta g^2 M, 
\label{su5RGEsg}
\end{eqnarray}
where $t = \ln(q^2/M_X^2)$, $q$ is the $\overline{MS}$ renormalization
scale, $g$ is the gauge coupling and the one loop beta function, $\beta =
2(S(R)-15)$, is determined by the sum over all the Dynkin indices of the
fields in the theory, $S(R)$. $\beta \! = \! -8$ for our SUSY SU(5) model,
which hosts the above mentioned Higgs sector plus $3(\underline{10} \oplus
\overline{\underline{5}})$ representations corresponding to 3 Standard
Model fermionic families (and superpartners). 
 
To render the fixed point structure explicit, from
(\ref{su5RGEsa}-\ref{su5RGEsg}) we form the following RG 
equations for
dimensionless ratios 
\begin{eqnarray}
16 \pi^2 \frac{d}{d t} \left( \frac{\lambda_1^2}{g^2} \right) &=& 6 g^2
\left( \frac{\lambda_1^2}{g^2} \right) \left[ \frac{189}{40} \left(
\frac{\lambda_1^2}{g^2} \right)  - 10 - \frac{\beta}{6} \right] \label{FP1}
\\ 
16 \pi^2 \frac{d}{d t} \left( \frac{m_3}{M \lambda_1} \right) &=& 9 g^2
\left[  \left( \frac{m_3}{M \lambda_1} \right) \left[ \left(
\frac{\lambda_1^2}{g^2} \right)  \frac{63}{20} - \frac{\beta}{9} \right] +
40 \right] \label{FP2} \\ 
16 \pi^2 \frac{d}{d t} \left( \frac{m_2^2}{M \mu} \right) &=&   g^2 \left[
- \beta \left( \frac{m_2^2}{M \mu} \right) +  \frac{63}{20} \left(
\frac{m_3}{M \lambda_1} \right) \left( \frac{\lambda_1^2}{g^2} \right) + 40
\right]~. \label{FP3} 
\end{eqnarray}
The right-hand side of this system of coupled equations vanishes for 
\begin{equation}
\left( \frac{\lambda_1^2}{g^2} \right)^* = \frac{40}{189} (10 + \beta / 6)
, \quad \left( \frac{m_3}{M \lambda_1} \right)^* = -6, \label{secFP} \quad
\left( \frac{m_2^2}{M \mu} \right)^* = - \frac{2}{3}. \label{thFP} 
\end{equation}
The fixed point described by Eq.~(\ref{thFP}) is a specific example of a
more general class of fixed points identified in Ref.~\cite{JJ}. 
By
linearizing (\ref{FP1}-\ref{FP3}) around the fixed point one easily finds
that it is infra-red stable when $\beta \! < \! 0$, as in the case of the
SUSY SU(5) model considered here. For $\beta \! > \! 0$, which can be
achieved by adding more matter to the model, one would have a saddle point.
Assuming $\delta_{3/2} \! \ll \! 1$, as implied by hierarchy arguments, we
may neglect the second order contribution of order $\delta_{3/2}^2$, and
$F$ is well approximated by 
\begin{equation} \label{fapp}
F\approx 3\delta_2-\frac{1}{3\lambda_1}\delta_3 =\frac{M}{\mu} \left[ 3
\frac{m_2^2}{M\mu} -\frac{1}{3} \frac{m_3}{M\lambda_1} \right]\;, 
\end{equation}
which is zero at the fixed point. Thus, starting at some scale $M^*$, {\it
e.g.} \/the Planck scale, and running to the GUT scale, the Higgs
parameters evolve towards values at the boundary ($F \! = \! 0$) between
the region of parameter space corresponding to unbroken SU(5) and the
region of parameter space corresponding to SU(5) breaking to
$G_{SM}$.

\begin{figure}[t]
\twographs{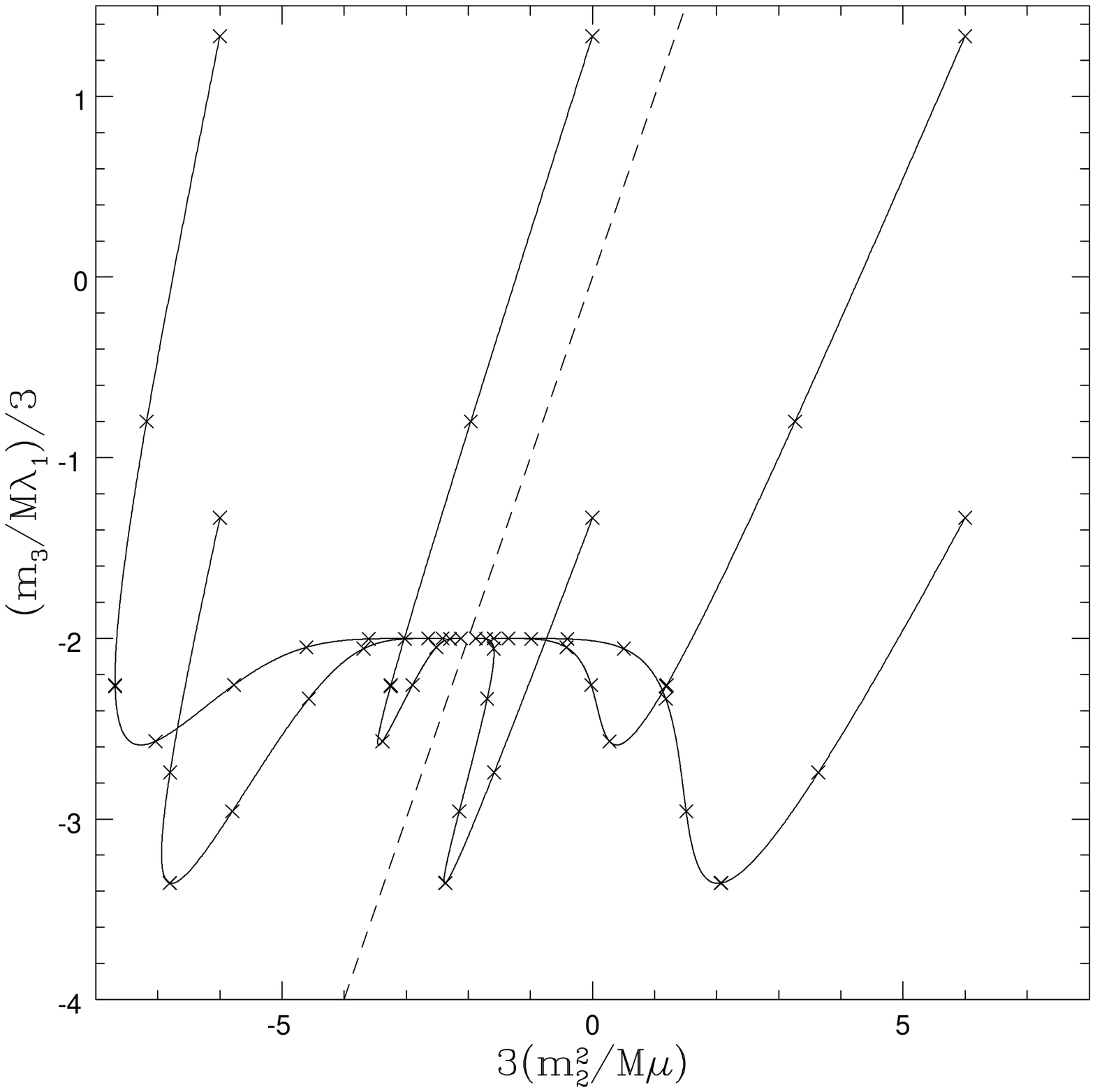}{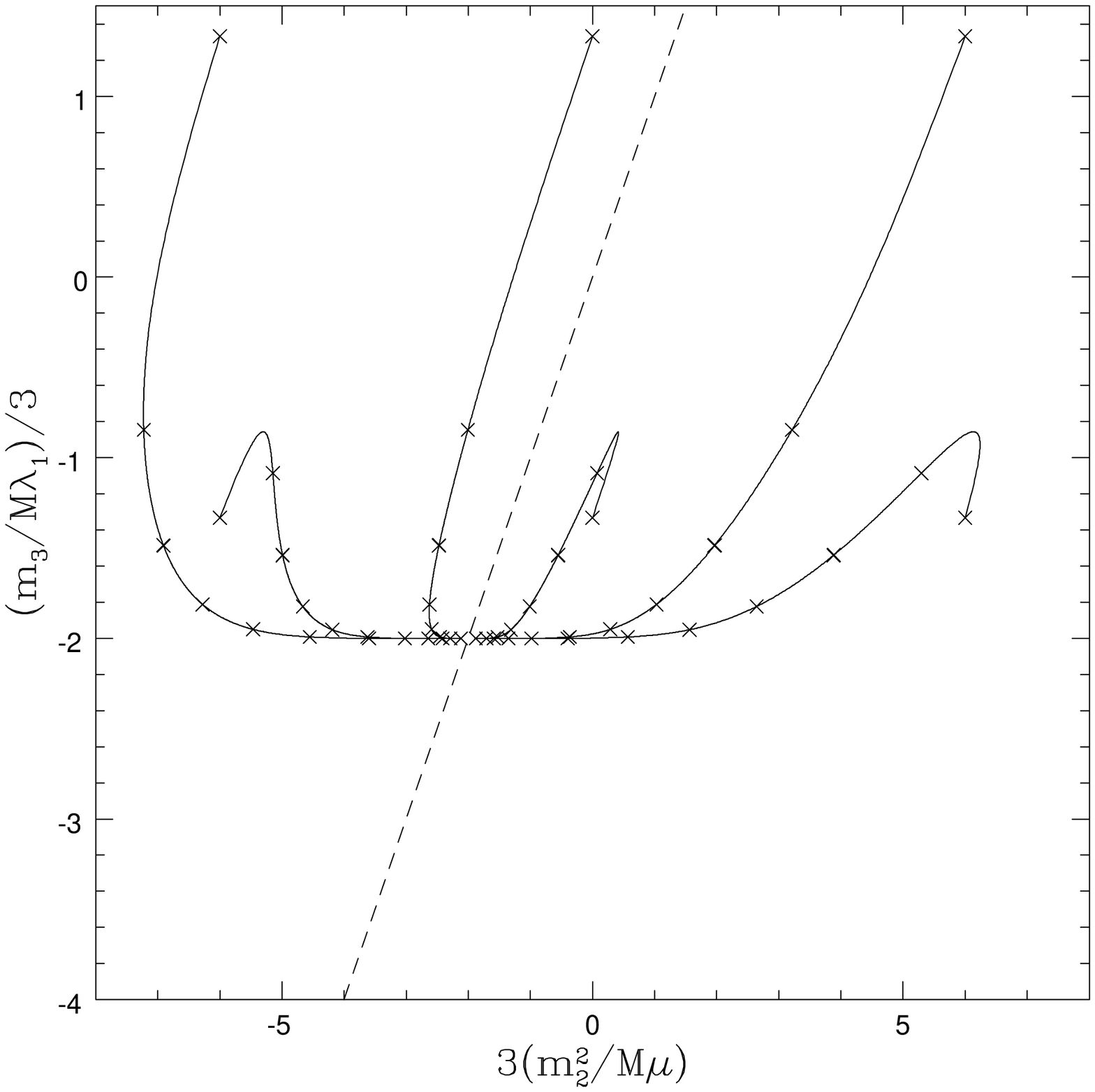}
\caption[]{\it RG flow between $M_P$ and $M_X$ of the soft SUSY-breaking
parameters in SUSY SU(5) with $\beta=-8$ for initial conditions with a)
$\lambda_1(M^*) \! = \! 0.3$ and b) $\lambda_1(M^*) \! = \! 2.0$. Every
decrease of the scale by a factor $10^{3/2}$ is marked on the flow.} 
\label{a}
\end{figure}
We have also studied our RG 
equations numerically for the parameter values $M^* \!
= \! M_P \!=\! 10^{19}$~GeV, $\mu (M^*) \! = \! 10^{16}$~GeV, $M(M^*)
\! = \! m_{3/2} (M^*) \! = \! 10^{3}$~GeV.
The gauge coupling is fixed
by $g^2 (M_X) \! = \! 8\pi/5$ to ensure consistency of SUSY SU(5)
unification of the Standard Model couplings with the low-energy experimental
data. The running parameters $M$ and $\mu$ evolve
slowly (they decrease by a factor of about $1/2$ between the Planck
and the GUT scale); consequently their ratio in (\ref{fapp}) does not change
sign. Hence, once the initial conditions are fixed, the sign of the
function $F$ depends on the relative magnitude of the combinations of
parameters $m_3/(3 M \lambda_1)$ and $3 m_2^2 / (M \mu)$. The flow of these
is depicted in Fig.\ref{a} for a small and a large initial value of
$\lambda_1 (M^*)$. The dashed line marks $3m_2^2/(M\mu)=m_3/(3M\lambda_1)$
where $F=0$. The region to the left of this line corresponds to the
breaking of SU(5) to $G_{SM}$ while the region to the right corresponds to
unbroken SU(5). 
%
%
%
%
For all the chosen initial values we have checked numerically that the
contribution of $\delta_{3/2}^2$ is indeed negligible over the whole range
of the running. The figure clearly displays the attracting fixed point;
however, the attraction is typically rather weak between the Planck scale
(first mark on the flow), and the GUT scale (third mark on the flow).
Interestingly, flows starting on the left (right) of the dashed line stay
on the left (right); therefore the flows never cross the boundary between
the region of parameter space corresponding to unbroken $SU(5)$ 
and the one
corresponding to $SU(5)$ breaking to $G_{SM}$. 
This general property implies that the running does not affect the amount
of tuning needed for the phenomenologically desirable scenario of SU(5)
breaking to $G_{SM}$, in the sense that the region of parameter space
supporting this scenario is mapped into itself by the RG 
flow\footnote{This would not be the case  
if there were significant contributions from $\delta_{3/2}^2$, 
since in that case 
the flow to the unbroken-SU(5) region is favored.}. 
We conclude
that, while it does not require any fine tuning, the scenario with $SU(5)$
breaking to the Standard Model is not a compelling prediction of the
infra-red RG structure of this SUSY $SU(5)$ GUT. 

\section{A non-SUSY SU(5) model}

Non-SUSY SU(5) with first step of SSB involving a $\underline{24}$ adjoint
Higgs was investigated in Ref.~\cite{SSBGUTs}.
Limiting again our analysis to the Higgs responsible for the first
step of SSB, we easily find that 
the Higgs potential can be written as \cite{orafrge} 
\begin{equation}
V(\phi) = -\frac{1}{2}m\mbox{Tr}\Phi^2-\frac{\sqrt{5}}{3!}f\mbox{Tr}\Phi^3+
\frac{5}{4!}g\mbox{Tr}\Phi^4+\frac{1}{4!}h(\mbox{Tr}\Phi^2)^2,
\label{su5V}
\end{equation}
where $\Phi$ is a traceless $5\times5$ matrix.

In the following, we will 
not consider the case  $f=g=0$,
since in this case the symmetry-breaking direction
is arbitrary (if $f=g=0$ the potential depends 
only on the norm of the
Higgs field). We emphasize that the classification in terms of the
cases $f=0,~f\neq0,~g=0,~g>0$, and $g<0$ is scale invariant \cite{orafrge}.

Let us begin assuming $f\neq0$ and $g=0$. In this case there is only one
possibility for the spontaneous symmetry breaking 
trajectory \cite{orafrge}: $SU(5)\rightarrow SU(4)\otimes U(1)$. 
At the transition
point the function $Q=3 m h/f^2$ vanishes. Since the derivative of $Q$ does
not vanish at this point, the changing of the scale can allow a change of
the symmetry from $SU(5)$ to $SU(4)\otimes U(1)$ and vice versa. $Q$ admits
a stationary point at $Q=189/880$, above which the symmetry is
$SU(4)\otimes U(1)$.

For $f=0$ and $g\neq0$ the breaking direction is independent of the scale and
the residual
symmetries are $SU(4)\otimes U(1)$ or $G_{SM}$
if $g<0$ or $g>0$ respectively.

\begin{figure}[tb]
%
%
%
\twographs{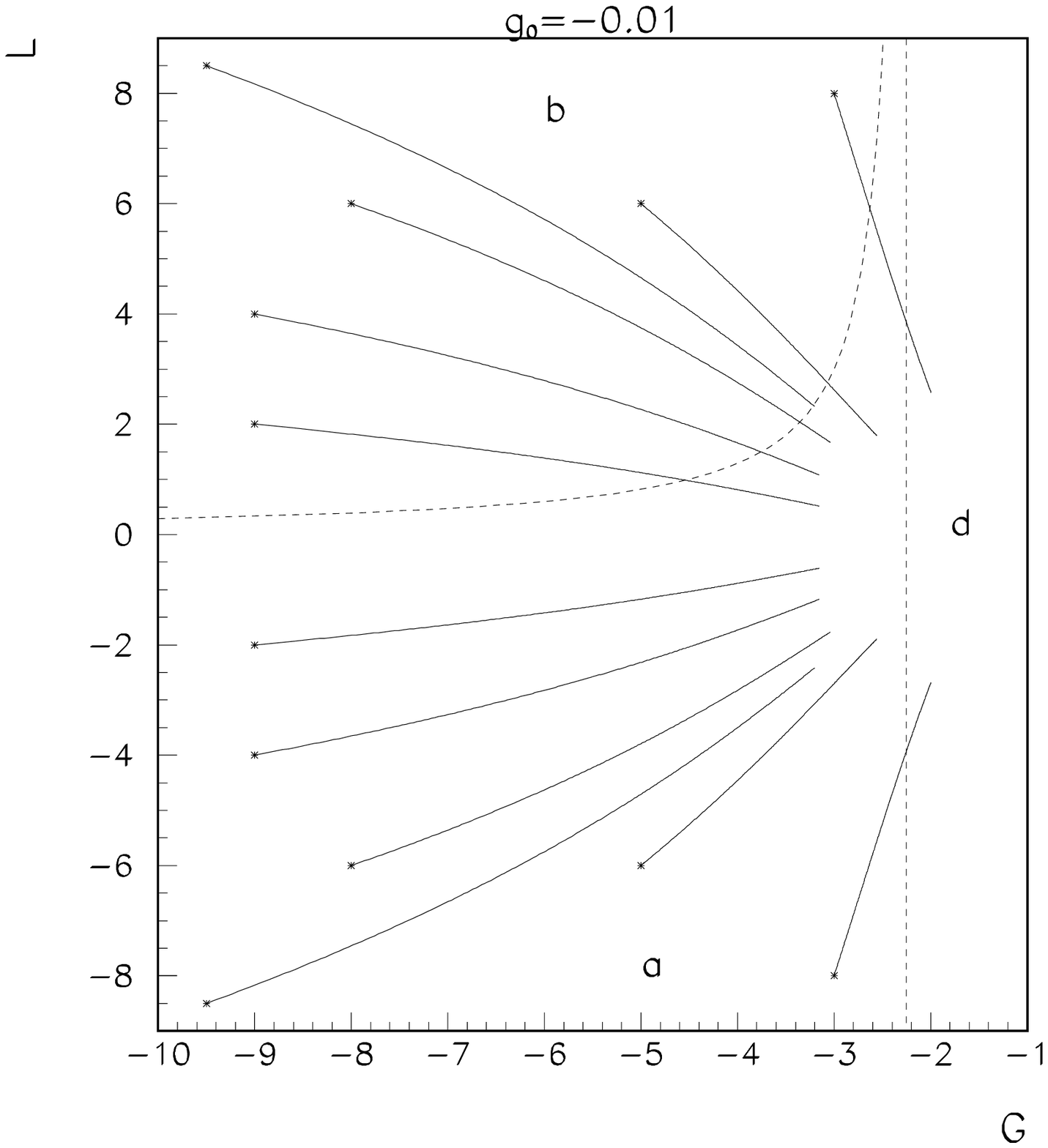}{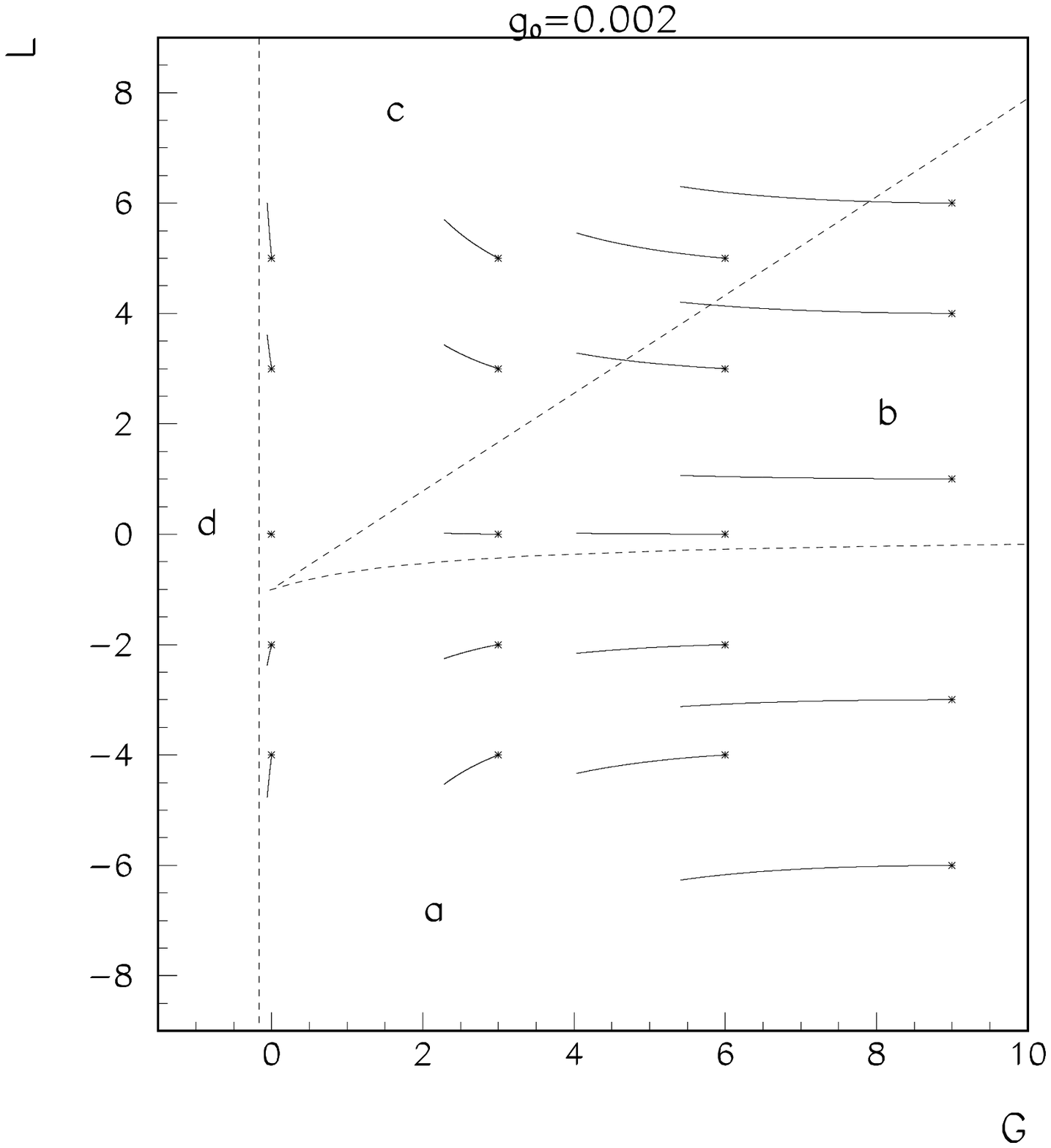}
\caption[]{\it RG flow between $M_P$ and $M_X$ of the
symmetry breaking
parameters in non-SUSY SU(5). Regions {\bf a}, {\bf b}, and {\bf c}
correspond to the symmetry groups SU(5), $SU(4) \otimes U(1)$, and $G_{SM}$
respectively. In region {\bf d} the
potential is unbounded from below.
Also notice that the initial condition (at $M_P$)
is marked on the flow.}
\label{f:su5}
\end{figure}

The case of interest
for our RG naturalness analysis is of course the
general case $f,~g\neq0$. The direction of symmetry breaking
is found to depend only on the following ratios:
\begin{equation}
L=\frac{3 m g}{f^2},~G=\frac{h+g}{g}. \label{bongo}
\end{equation}
The corresponding RG 
equations are:
\begin{eqnarray}
\frac{d L}{dt} &=& g\left( \frac{52}{75} L G-\frac{6386}{975}L -\frac{63}{100} 
\right), \label{eq:mg} \\
\frac{d G}{dt} &=& g\left( \frac{56}{25} G^2+\frac{652}{195}G+
\frac{28558}{12675} \right), \label{eq:gg} \\
\frac{d g}{dt}  &=& g^2 \left( \frac{8}{25}G +\frac{8}{15} \right). 
\label{eq:gp} 
\end{eqnarray}
The RG flows, for $L$ and $G$, are depicted for the two
cases $g_0\equiv g(M^*)=0.002$ and $g_0=-0.01$,
with $M^*=M_P$.
These two cases\footnote{The very small values of $|g_0|$ shown 
in Fig.\ref{f:su5}
have been chosen because they allow to illustrate more clearly 
the qualitative structure of the RG running, which
only depends on the sign of $g_0$.
In any case one should in principle only consider $g_0 \ll 1$
so that our one-loop equations are reliable.} 
are representative of the two typical 
situations $g_0<0$ and $g_0>0$.
In both figures 
the dominant qualitative behavior is the one associated
to the fixed point $L=0, G \rightarrow \infty$.
This fixed point can be seen directly from
the Eqs.~(\ref{eq:mg})-(\ref{eq:gp}) by observing that
for $g=0$ the right-hand side of 
this system of coupled equations vanishes.
The limit $L=0, G \rightarrow \infty$ follows
from  $g=0$ as a result of
Eq.~(\ref{bongo}).
As seen from Fig.\ref{f:su5}
the fixed point 
is infra-red stable when $g_0 \! < \! 0$, whereas for $g_0 \! > \! 0$
it gives a saddle point.

Let us now discuss the implications for symmetry breaking 
of this RG analysis, starting with the case $g_0=-0.01$. 
Here the $G$ values are
restricted to the $(-\infty,-9/4)$ interval, otherwise the potential
becomes unbounded from below. The vertical dashed line 
in Fig.\ref{f:su5}a corresponds to
$G=-9/4$ and separates this region ({\bf d}) from the other ones. 
If $L>-1/(1+4G/9)$ the symmetry group is given by $ SU(4)\otimes
U(1)$ (region {\bf b}), while in the opposite case there is no symmetry
breaking at all (region {\bf a}). From the figure we see that 
if $L<-1/(1+4G/9)$ 
the curves remain in the $SU(5)$-symmetric
region, but in some cases they end up in {\bf d}-region. On the other hand, 
we observe that when the flows start with $L>-1/(1+4G/9)$ 
({\it i.e.} they start from the {\bf b}-region)
they quickly come out
of the region of $SU(4)\otimes U(1)$ symmetry and end up in the
$SU(5)$-symmetric region.
We conclude that for $g_0<0$
the breaking of $SU(5)$ 
appears to be quite unnatural in light of 
the results of the RG analysis,
even though the non-dynamical part of the analysis
presented us with a symmetry-breaking ({\bf b}) region 
of size roughly comparable 
to the one of the symmetric-preserving ({\bf a}) region.

In the case $g_0>0$ the $(L,G)$ plane is divided into three regions, {\bf
a}, {\bf b}, and {\bf c}, which correspond to the symmetries $SU(5)$,
$SU(4)\otimes U(1)$, and $G_{SM}$ respectively. Note
that for $G<-1/6$ the potential is unbounded from below and, once
again, in Fig.\ref{f:su5}b
the vertical dashed line at $G=-1/6$ separates the {\bf d}-region
from the other ones. 
Importantly, the curves that originate in the {\bf c} region 
remain in that part of the plane, whereas the flows with
starting point in the {\bf b}-region
rather quickly (depending on how
close the starting point is to the line $L=-1 + 8G/9$)
end up in the {\bf c}-region.
This implies that for $g_0>0$
the breaking of $SU(5)$ to $SU(4)\otimes U(1)$ 
is quite unnatural, whereas
the breaking of $SU(5)$ to $G_{SM}$
does not require any fine-tuning. Actually,
as a result of the fact that some of the flows starting in the
{\bf b}-region end up in the {\bf c}-region,
the RG flow maps a larger portion of parameter space
at the scale $M^*$ into the {\bf c}-region of parameter space,
which corresponds to the breaking of $SU(5)$ 
to $G_{SM}$.
However, this portion of Higgs parameter space
that corresponds to the ``phenomenologically
reasonable" symmetry breaking to $G_{SM}$
is not a general attractor for the RG flow; in fact,
flows starting in the {\bf a}-region remain in that region,
so that the case of unbroken $SU(5)$ is also quite
consistent with our RG analysis.
Still, as the present article is searching for qualitative 
structures that might in general characterize
this type of RG analyses
it is interesting to notice that even in nonSUSY $SU(5)$
there are scenarios in which
the regions of parameter space supporting symmetry breaking
are stable with respect to the ones supporting unbroken $SU(5)$; in fact,
there is no flow across the $L = -1 / (1+ 4G/9)$ curve.

For completeness, in closing this section let us comment
on the fact that in Fig.\ref{f:su5}
there is an approximate symmetry
with respect to the $L\rightarrow -L$ exchange. 
This reflects the fact that the Eqs.~(\ref{eq:mg})-(\ref{eq:gp}) 
are invariant with respect to the $L\rightarrow -L$ exchange when 
$L$ and $G$ are large enough that
one can ignore the factor $-63/100$.

\section{A non-SUSY SO(10) model}

Non-SUSY SO(10) with first step of SSB involving a $\underline{54}$ Higgs
was investigated in Refs.~\cite{MandR,so1054,lone}. 
Besides the $\underline{54}$,
the Higgs sector of the model also includes
$\underline{10} + \underline{126} + \underline{126}^*$ Higgs plus $3 \times
\underline{16}$ representations containing the SM fermions and 3
right-handed neutrinos. Again, due to the hierarchy among the scales
involved and the fact that we are here concerned only with the first step
of SSB, we only consider the Higgs potential involving the Higgs
responsible for the first step of SSB. The most general Higgs potential
constructed from a $\phi \sim \underline{54}$ Higgs may be written
as~\cite{MandR} 
\begin{equation}
V(\phi) = \lambda \mbox{Tr} \left[ ( \phi^2 - \frac{1}{10} \mbox{Tr}
(\phi^2) + a \phi )^2 \right] - \lambda b^2 \mbox{Tr} (\phi^2) +
\frac{\mu}{60} \left[ \mbox{Tr} (\phi )^2 \right]^2. \label{so10V}
\end{equation}
and the $SO(10)$-invariance may be spontaneously broken
to SO(m)$\otimes$SO(10-m) with $m=1,2,3,4,5$ depending upon 
the combinations of
parameters $b^2 / a^2$ and $\mu / \lambda$. 
Of course, the ``phenomenologically
reasonable" region of Higgs parameter space is the one that
supports symmetry breaking of $SO(10)$ 
to $SO(6) \otimes SO(4)$, which contains the Standard Model gauge group.
In the following we shall denote 
with {\bf $\alpha$}, {\bf $\beta$}, {\bf $\gamma$}, {\bf $\delta$},
and {\bf $\sigma$} respectively the regions of Higgs parameter
space that correspond to the breaking of $SO(10)$ to $SO(5) \otimes SO(5)$, 
$SO(6) \otimes SO(4)$, 
$SO(7) \otimes SO(3)$, 
$SO(8) \otimes SO(2)$ and $SO(9)$.
As it will not be visible in our figures we will simply refer
to the region of parameter space
corresponding to unbroken $SO(10)$ as the $SO(10)$-region.

The radiative corrections to the Higgs potential were examined in
Ref.~\cite{MandR} and regions of parameter space yielding ``stable minima''
(in the specific sense of \cite{MandR}) were searched for. The objective
of our approach is somewhat different in that we investigate the running
between some high scale $M^*$ and the GUT scale $M_X$ in order to find the
most natural {\it intermediate symmetry group}, i.e. the most natural group
of residual symmetry after the first (GUT-scale) step of SSB. The relevant
RG 
equations are~\cite{MandR} 
\begin{eqnarray}
16 \pi^2 \frac{d \lambda}{d t} &=& 48  \lambda \left(
\frac{\mu}{60}-\frac{\lambda}{10}  \right) + \frac{508}{35} \lambda^2 +
\frac{15}{2} g^4 - 60 g^2 \lambda \nonumber \\ 
16 \pi^2 \frac{d }{d t} \left( \frac{b^2}{a^2} \right) &=& \lambda \left[
\left( \frac{b^2}{a^2}  - 1 \right) \left( - \frac{342}{5} + \frac{56}{15}
\frac{\mu}{\lambda} + \frac{15}{2} \frac{g^4}{\lambda^2} \right) -
\frac{504}{5} \right] \nonumber \\ 
16 \pi^2 \frac{d }{d t}\left( \frac{\mu}{\lambda} \right) &=& \lambda
\left[ \frac{1264}{5} - \frac{6}{5} \frac{\mu}{\lambda} + \frac{10}{3}
\left( \frac{\mu}{\lambda} \right)^2 + \frac{15}{2} \frac{g^4}{\lambda^2}
\left( 24 - \frac{\mu}{\lambda} \right) \right] \nonumber \\ 
16 \pi^2 \frac{d g^2}{d t} &=& -\frac{70}{3} g^4. \label{so10RGEs} 
\end{eqnarray}
We have been unable to find combinations of parameters allowing the
identification of a fixed point, and therefore we proceed directly to a
numerical approach. 

\begin{figure}[tb]
%
%
%
\twographs{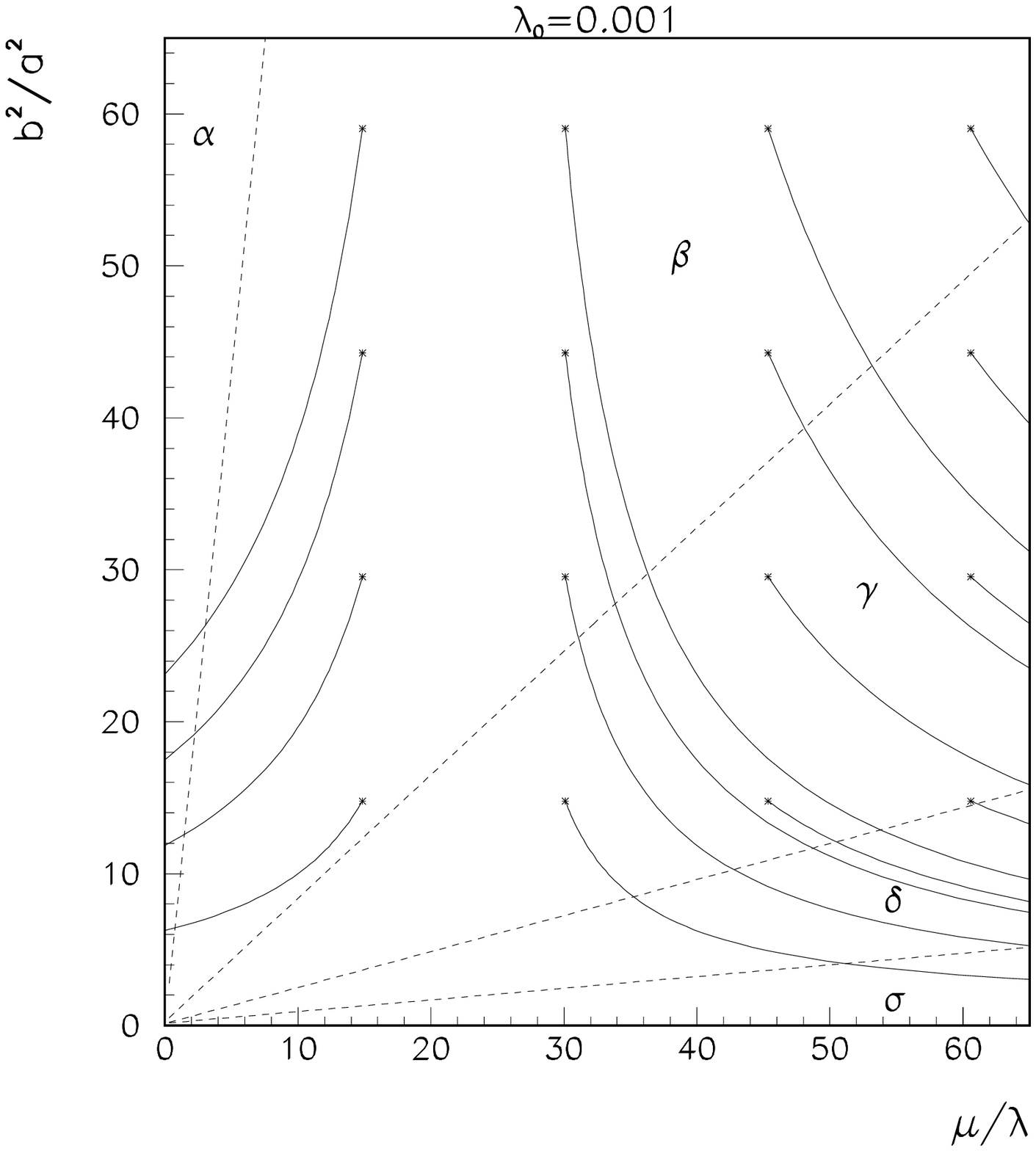}{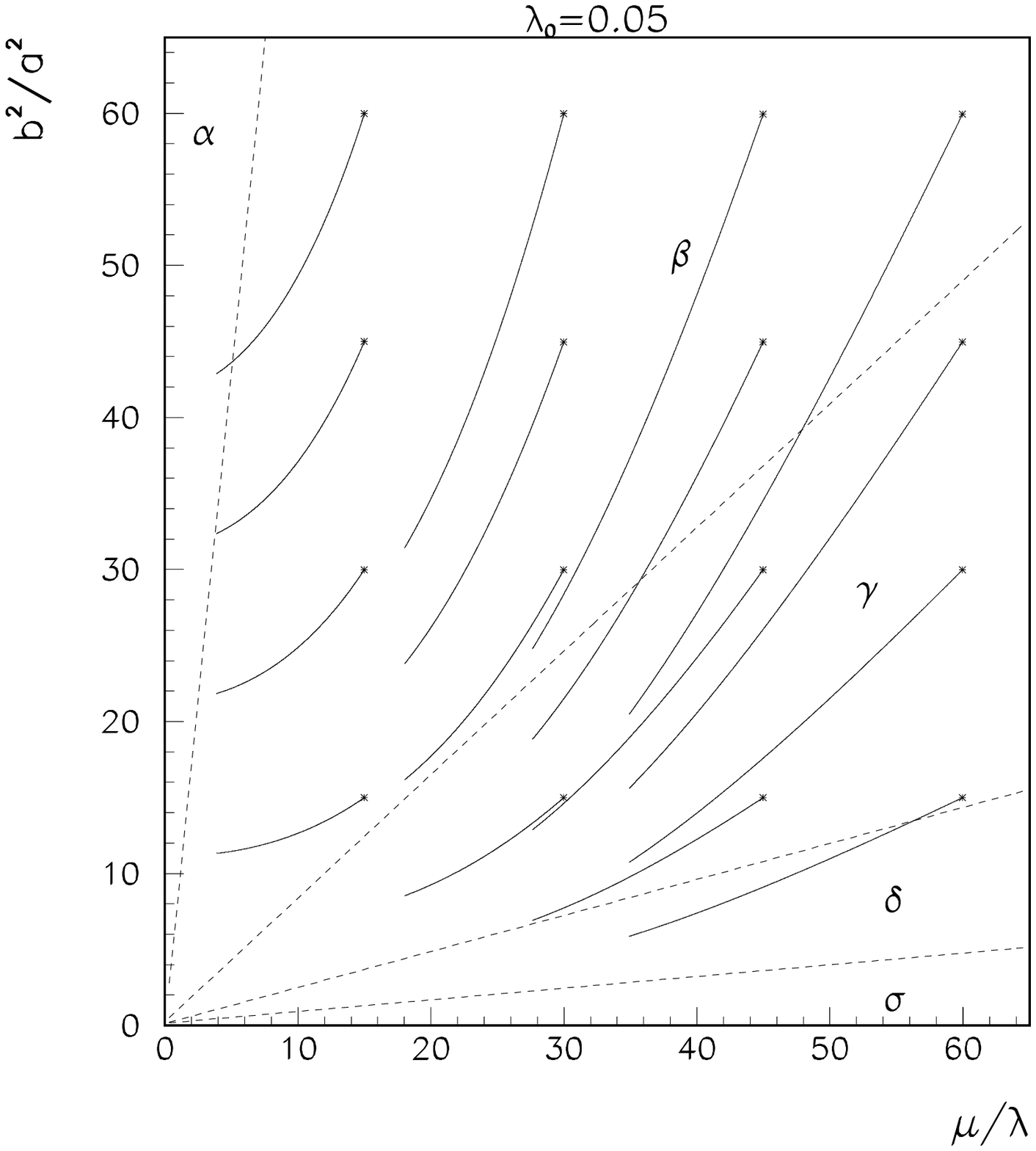}
\caption[]{\label{c} \it RG flow between $M_P$ and $M_X$ of the 
symmetry-breaking parameters in non-SUSY SO(10).
The regions {\bf $\alpha$}, {\bf $\beta$}, {\bf $\gamma$}, {\bf $\delta$},
and {\bf $\sigma$}
correspond to the breaking of $SO(10)$ to $SO(5) \otimes SO(5)$, 
$SO(6) \otimes SO(4)$, 
$SO(7) \otimes SO(3)$, 
$SO(8) \otimes SO(2)$ and $SO(9)$ respectively. 
In the parametrization here adopted the region of parameter space
corresponding to unbroken $SO(10)$ is located below 
the {\bf $\sigma$}-region
but it is too small to be seen in figure.
Also notice that the initial condition (at $M_P$)
is marked on the flow.}
\end{figure}

Here we work with $g^2(M_X)=(4\pi)/42$. Fig.\ref{c} shows the
renormalization group flow for the parameters relevant for symmetry
breaking, $\mu/\lambda$ and $b^2/a^2$, for various choices of initial
conditions at $M^*=M_P$. 
The figures clearly show that there are
regions of attraction 
for the flow of the couplings. In the case of
large initial values of $\lambda$ there is a strong attraction to the
SO(5)$\times$SO(5) region of parameter space. With decreasing
$\lambda(M^*)$ this attraction becomes weaker until, at about
$\lambda(M^*)\approx 0.03$, a new feature emerges:
a dividing line which is not crossed. Everything to the right of this line is
attracted to the $SO(9)$- and $SO(10)$-invariant regions, while everything 
to the left of this line is still attracted toward 
the $SO(5) \otimes SO(5)$-invariant region.
The region of parameter space leading to ``phenomenologically
reasonable" symmetry breaking, {\it i.e.} the one
corresponding to intermediate symmetry $SO(6) \otimes SO(4)$,
does not appear to require large
fine tuning, although the RG flow maps a smaller portion of parameter space
at the scale $M^*$ into the ``phenomenologically reasonable" region of
parameter space. 

\section{Cosmology and Supercosmology}

In the previous sections we have concentrated on a zero-temperature
analysis. However, clearly an important constraint on GUTs is the
consistency with a working cosmological scenario, and checking this
consistency requires in general a finite-temperature analysis. 
While we
postpone this type of analysis to future work, in this section we 
make a few observations concerning the cosmological relevance of
studies of the type reported in the previous sections.

We observe that there is a substantial difference between
the SUSY and the non-SUSY cases. In non-SUSY GUTs
the potential-energy difference between the absolute
minimum of the Higgs potential and other minima is of order the GUT scale.
This implies that at temperatures of order $M_X$, {\it i.e.} once the
temperature is low enough that the features of the zero-temperature
effective potential become relevant (at high temperatures thermal effects
dress the potential in such a way that the absolute minimum
is necessarily symmetric~\cite{gacsnr}), 
the universe rapidly reaches the vacuum corresponding to the
absolute minimum of the zero-temperature effective potential. Therefore for
non-SUSY GUTs an analysis of the type presented in the previous sections
is directly relevant for the understanding of certain cosmological issues,
such as the the selection of the presently observed vacuum. 

An important factor affecting the corresponding analysis of SUSY GUTs
is the near degeneracy (up to SUSY-breaking terms) of several minima,
which we mentioned in Sec.2. The energy difference between the absolute
minimum and the other minima is of order SUSY-breaking terms, and therefore
much smaller than the GUT scale. 
As a result, 
at least within a
perturbative analysis, 
one finds that even when the temperature becomes low enough
for the features of the zero-temperature effective potential to become
relevant, the universe does not rapidly reach the vacuum corresponding to
the absolute minimum\footnote{A recent 
analysis by Abel and Savoy~\cite{as} of charge and color breaking (CCB) 
minima in the MSSM 
showed that a global CCB minimum is neither
necessary nor sufficient for symmetry breaking. Abel and Savoy then
concentrated on a sufficient condition for no symmetry breaking, namely that
the radiatively corrected potential not contain a {\em local} CCB minimum.
Analyses such as the one we perform in the present article
does not provide insight in issues
relevant for cosmological studies of the type reported 
in Ref.~\cite{as}, since we focus
on the identification of the global
minimum in the zero temperature potential.}
of the zero-temperature effective
potential \cite{longtime}. Actually, quick estimates within ordinary
perturbative approaches are sufficient to show that the time needed for the
transition to the true vacuum should be expected to be longer than the
lifetime of the universe \cite{longtime}. 

One way to obtain working supercosmology \cite{supercosmo,strongcoupling}
scenarios is to advocate strong-coupling \cite{strongcoupling} thermal
effects, which are indeed at work in the SUSY case \cite{kapu}. The
investigation of these issues requires a careful (and very delicate)
thermal analysis which goes beyond the scope of this article. However, it
should be noticed that the type of analysis given in Sec.2 is not very
relevant to this type of supercosmological issues. 

A more conventional (but {\it ad hoc}) way to obtain working
supercosmological scenarios is based \cite{meltemp} on fine tuning of the
parameters of the Higgs potential. One scales down the entire
superpotential, so that the height of the potential barrier between
competing vacua becomes of the same order of their energy difference, while
keeping fixed the mass of the gauge bosons mediating proton decay. For
example in the minimal SUSY SU(5) GUT discussed in Sec.2 one would 
divide \cite{meltemp} both $\lambda$ 
and $\mu$ by a common (so that the ratio $\mu / \lambda$ giving 
mass to the gauge bosons mediating proton decay remains
unchanged) large factor, of order $10^{12}$. Analyses of the type advocated
in the present paper could be relevant for this supercosmology scenario;
one can in fact check the level of fine tuning at the Planck scale needed
to have a $10^{-12}$ fine tuning at the GUT scale. We find that the
fine-tuned values of $\lambda$ and $\mu$ are so far from the region of
attraction of the fixed point that the RG running between $M_P$ and $M_X$
is not substantial: a fine tuning of $10^{-13}$ is required at $M_P$ in
order to obtain a $10^{-12}$ fine tuning at the GUT scale. 

\section{Closing Remarks}

The three GUTs that were considered in the previous sections
have allowed us to illustrate
various possibilities for the outcome of RG naturalness
analyses, although a common property of these illustrative examples
is that one does not find a compelling SSB scenario compatible
with the low-energy phenomenology of the Standard Model.
This could be interpreted positively, since these
three GUTs are already known to give rather unsatisfactory
predictions for low-energy observables,\footnote{For example,
the minimal SUSY $SU(5)$ GUT does not
address the question of doublet-triplet splitting~\cite{doubtrip},
while the nonSUSY $SU(5)$ and $SO(10)$ GUTs here considered
are not consistent with the available data on proton stability
(see, {\it e.g.}, Ref.~\cite{lone}).}
even when their SSB pattern
is tuned (by tuning the parameters of the Higgs potential)
to be group-theoretically consistent with the Standard Model.
One is tempted to hope that applying the same naturalness criterion
to the GUTs that are already known to have good
low-energy phenomenology it would be possible to find one 
for which even the SSB pattern appeared compelling ({\it i.e.}
such that the values of the parameters of the Higgs
potential that correspond to this SSB pattern could
be obtained via RG running from rather
generic input parameters at the scale $M^*$).
Models with enough structure to be consistent with
low-energy phenomenology will require 
rather complicated RG naturalness analyses
from the technical point of view,
but the conceptual steps are of course just
the ones discussed here.

Among the properties illustrated by our three simple illustrative
examples, particularly significant is the fact that
in all the three cases we find that the RG naturalness
criterion appears to favor
spontaneous-symmetry-breaking $G \rightarrow H$ scenarios
such that the residual-symmetry group $H$
is the ``smallest'' (the one with the smallest number of generators)
of the maximal little groups
of the group $G$ which is being broken.
In fact, both in the SUSY and in the nonSUSY cases with $SU(5)$
grand unification group 
the breaking of $SU(5)$ to $SU(4)\otimes U(1)$ 
was quite unnatural in all scenarios considered,
whereas under appropriate conditions 
the breaking of $SU(5)$ to $G_{SM}$
(which is the smallest
of the maximal little groups of $SU(5)$)
appeared to be as natural as the possibility that
$SU(5)$ would remain unbroken.
Similarly, in our analysis of the first step of SSB 
of a nonSUSY $SO(10)$ model we found that
under certain conditions 
the breaking of $SO(10)$ to $SO(5) \otimes SO(5)$
(which is the smallest
of the maximal little groups of $SO(10)$)
appeared to be as natural as the possibility that
$SO(10)$ would remain unbroken, while all other possibilities
appeared to be quite unnatural from the RG viewpoint
(which is quite unfortunate since $SO(5) \otimes SO(5)$
does not contain the Standard Model gauge group).
It appears that the RG running favors either 
the maximal preservation of symmetries (unbroken $G$)
or the maximal breaking of symmetries
(which in a specific sense corresponds to the breaking
of $G$ to its smallest maximal little group).
It would be quite interesting to analyze from the point of view
of RG naturalness some of the known counterexamples 
to Michel's conjecture, {\it i.e.} GUTs whose Higgs potential 
have enough structure to allow breaking to a non-maximal
little group (see, {\it e.g.}, the example 
discussed in Ref.~\cite{lone}).

Our findings also suggest
(as it was to be expected) that there are many technical 
differences between the RG naturalness analysis of a SUSY GUT 
and the corresponding analysis of a nonSUSY GUT.
For example, the simplifications associated to SUSY
allow one to find analytically a fixed point in the case
of SUSY $SU(5)$.

While we focused on the first step of SSB in our discussion and examples,
the criterion of RG naturalness can obviously be applied
to any of the steps of a SSB pattern.
Actually, in some cases the criterion (at least as applied
in the present article) might be least meaningful
when considering the first step of SSB.
In fact, it appears plausible that only two or three 
orders of magnitude would
separate the scale $M^*$ from $M_X$, in which case
omitting non-renormalizable terms, as we have done above, might be
unjustified. Several orders of magnitude instead should separate 
the GUT scale $M_X$ from the scale $M_{II}$ of the second step 
of SSB, which could be the electroweak scale or (in models
with an intermediate symmetry group \cite{lone})
a scale of order $10^9 \! \sim \! 10^{12}$ GeV.
Therefore, most practical applications of the ideas here presented
might end up being found in studies of the running between
$M_X$ and $M_{II}$, for which non-renormalizable terms
can be safely ignored, rather than the running between
$M^*$ and $M_X$ which we have here considered 
for illustrative purposes.

Perhaps the most robust observation one can make based
on the results here reported is that
the conventional tests of the naturalness of a GUT
are quite inadequate.
According to these conventional naturalness
tests one should disregard
GUTs in which a fine tuning of the Higgs parameters is needed
in order to realize a phenomenologically acceptable SSB pattern.
However, viewing the GUTs as effective low-energy
descriptions of a more fundamental theory
one would like to check whether the
phenomenological SSB pattern
corresponds to fine tuning
of the Higgs parameters at the scale $M^*$.
Some of our results indicate that it is not uncommon
that a scenario requiring no fine tuning
of the Higgs parameters at $M^*$
might correspond via the RG running (for example in presence of
an appropriate infra-red fixed point) to a narrow
region (apparent fine tuning) of the Higgs parameter space
at $M_X$, where the first step of SSB is decided.
(Of course, corresponding statements should apply to the other steps
of the SSB pattern.)
Our analysis also provided examples of the opposite, {\it i.e.} 
a SSB pattern 
that in a ``conventional naturalness test'' would appear
to correspond to a significant
portion of the Higgs parameter space
actually requires some level of fine tuning at $M^*$,
since the corresponding portion of Higgs parameter space
is ``disfavored" by the RG running.

\section*{Acknowledgments}
We have greatly benefited from conversations with G.~Ross, 
which we happily 
acknowledge. One of us (G.A.-C.) also thanks
K.~Tamvakis for useful discussions.
The work of G.A.-C. was supported by a grant of
the Swiss National Science
Foundation, by a European Union TMR-network grant
(no.~FMRXCT960045), and by an OFES grant (no.~95.0856) 
of the Swiss Federal Office for Education and Science.

\end{document}